\documentclass[letterpaper,10pt,twocolumn,superscriptaddress,aps,prl]{revtex4-1}
\usepackage{amsfonts,amssymb,amsmath,amsthm,graphicx}
\usepackage{url}
\usepackage{comment}
\usepackage{siunitx}
\usepackage[colorlinks=true, citecolor=blue,urlcolor=blue,linkcolor=blue,filecolor=black]{hyperref}

\begin{document}

\title{Tools for the Performance Optimization of Single-Photon\\ Quantum Key Distribution}

\author{Timm Kupko}
\affiliation{Institut f\"ur Festk\"orperphysik, Technische Universit\"at Berlin, 10623 Berlin, Germany}
 
\author{Martin v. Helversen}
\affiliation{Institut f\"ur Festk\"orperphysik, Technische Universit\"at Berlin, 10623 Berlin, Germany}

\author{Lucas Rickert}
\affiliation{Institut f\"ur Festk\"orperphysik, Technische Universit\"at Berlin, 10623 Berlin, Germany}

\author{Jan-Hindrik Schulze}
\affiliation{Institut f\"ur Festk\"orperphysik, Technische Universit\"at Berlin, 10623 Berlin, Germany}

\author{Andr\'{e} Strittmatter}
\affiliation{Institut f\"ur Festk\"orperphysik, Technische Universit\"at Berlin, 10623 Berlin, Germany}
\affiliation{Present address: Institut f\"ur Experimentelle Physik, Otto-von-Guericke Universit\"at Magdeburg, 39106 Magdeburg, Germany}

\author{Manuel Gschrey}
\affiliation{Institut f\"ur Festk\"orperphysik, Technische Universit\"at Berlin, 10623 Berlin, Germany}

\author{Sven Rodt}
\affiliation{Institut f\"ur Festk\"orperphysik, Technische Universit\"at Berlin, 10623 Berlin, Germany}

\author{Stephan Reitzenstein}
\affiliation{Institut f\"ur Festk\"orperphysik, Technische Universit\"at Berlin, 10623 Berlin, Germany}

\author{Tobias Heindel}
\email[Corresponding author: ]{tobias.heindel@tu-berlin.de}
\affiliation{Institut f\"ur Festk\"orperphysik, Technische Universit\"at Berlin, 10623 Berlin, Germany}

%%%%%%%%%%%%%%%%%%%%%%%%%%%%%%%%%%%%%%%%%%%%%%%%%%%%%%%%%%%%%%%%%%%%%%%%%%%

%%%%%%%%%%%%%%%%%%%%%%%%%%%%%%%%%%%%%%%%%%%%%%%%%%%%%%%%%%%%%%%%%%%%%%%%%%%

\begin{abstract}
Quantum light sources emitting triggered single photons or entangled photon pairs have the potential to boost the performance of quantum key distribution (QKD) systems. Proof-of-principle experiments affirmed these prospects, but further efforts are necessary to push this field beyond its current status. In this work, we show that temporal filtering of single-photon pulses enables a performance optimization of QKD systems implemented with realistic quantum light sources, both in experiment and simulations. To this end, we analyze the influence of temporal filtering of sub-Poissonian single-photon pulses on the expected secret key fraction, the quantum bit error ratio, and the tolerable channel losses. For this purpose, we developed a basic QKD testbed comprising a triggered solid-state single-photon source and a receiver module designed for four-state polarization coding via the BB84 protocol. Furthermore, we demonstrate real-time security monitoring by analyzing the photon statistics, in terms of $g^{(2)}(0)$, inside the quantum channel by correlating the photon flux recorded at the four ports of our receiver. Our findings are useful for the certification of QKD and can be applied and further extended for the optimization of various implementations of quantum communication based on sub-Poissonian quantum light sources, including measurement-device-independent schemes of QKD as well as quantum repeaters. Our work represents an important contribution towards the development of QKD-secured communication networks based on quantum light sources.
\end{abstract}

\maketitle

\section{Introduction}
Privacy in communication is an increasingly important challenge in our information-driven society \cite{Acin2018}.
The concepts gathered in the field of quantum communication \cite{Gisin2002,Lo2014,Diamanti2016} represent solutions to this challenge and enable information theoretical secure communication. Quantum key distribution (QKD) for instance enables the tap-proof encryption of data, by exploiting quantum properties of light \cite{Bennett1984,Ekert1991}. The respective quantum light sources ideally required for QKD, however, had been impossible to fabricate with sufficient brightness and quality for a long time. Most implementations of QKD are therefore still implemented with weak coherent pulses (WCPs), i.e. attenuated lasers, requiring so-called decoy-state protocols \cite{Wang2005, Lo2005}. During the last decade, however, tremendous progress has been made in the fabrication of quantum light sources. Single-photon sources (SPSs) based on epitaxial semiconductor quantum dots (QDs) nowadays can be triggered at GHz clock rates under pulsed-optical \cite{Schlehahn2015a} and -electrical \cite{Hargart2013,Schlehahn2016} excitation, feature high degrees of photon indistinguishability \cite{Somaschi2016,Wang2016}, large photon extraction efficiencies \cite{Heindel2010,Schlehahn2016,Wang2019}, and to date achieve the highest single-photon purity in terms of $g^{(2)}(0)$ compared to any other single-photon emitter \cite{Schweickert2018,Hanschke2018}. The development of deterministic fabrication technologies had particular large impact for these developments, as summarized in a recent review article in \cite{Rodt2020}. Despite this immense progress, only few proof-of-concept QKD experiments have been reported based on optically \cite{Waks2002,Beveratos2002,Alleaume2004,Intallura2007,Collins2010,Takemoto2010} and electrically \cite{Heindel2012,Rau2014} operated single-photon sources. These experiments affirmed the potential sub-Poissonian light sources offer for QKD. To push the field of sub-Poissonian QKD to a new level, however, further efforts need to be undertaken. In particular practical methods for the security analysis and certification as well as measures to improve the performance of QKD systems for a given quantum light source need to be developed. While Waks et al. discussed security aspects of QKD with sub-Poissonian light sources from a theoretical viewpoint \cite{Waks2002a}, experimental studies on this important topic are still missing.

In this work, we perform a detailed analysis on the influence of temporal filtering of single-photon pulses on the performance of QKD systems implemented with sub-Poissonian light sources. For this purpose we set up a basic QKD testbed comprising a QD-based SPS and a receiver module designed for four-state polarization coding via the BB84 protocol. Using this Bob module in combination with our SPS, we determine the sifted key fraction, the quantum bit error ratio (QBER) caused by the receiver, and the $g^{(2)}(0)$ of the single-photon pulses inside the quantum channel, to finally extract the secure key rate expected in full implementations of QKD. As the temporal filtering of single-photon pulses differently affects these parameters, a performance optimization of QKD systems implemented with quantum light source is possible. We show that optimal performance for a given SPS can be achieved by carefully setting Bob's acceptance time windows, depending on the pulse shape and noise level. This can be either used to maximize the secure key rate for a given channel loss or to extend the maximally tolerable loss, i.e. the achievable communication distance. In addition, we demonstrate real-time security monitoring by analyzing the suppression of multi-photon emission events, i.e. $g^{(2)}(0)$ of the single-photon pulses inside the quantum channel during key generation. Finally we generalize our findings by employing simulations with synthetic pulse shapes, providing predictions for different SPSs and detectors. We consider the results presented in this work an important contribution towards the development of QKD-secured communication networks based on quantum light sources. Importantly, our approach can be easily applied and further extended for the optimization of any implementation of quantum communication based on sub-Poissonian quantum light sources.

\section{Results}
\label{sec:results}

\subsection*{QKD testbed}
\begin{figure*}
	\centering
	\includegraphics[width=0.95\linewidth]{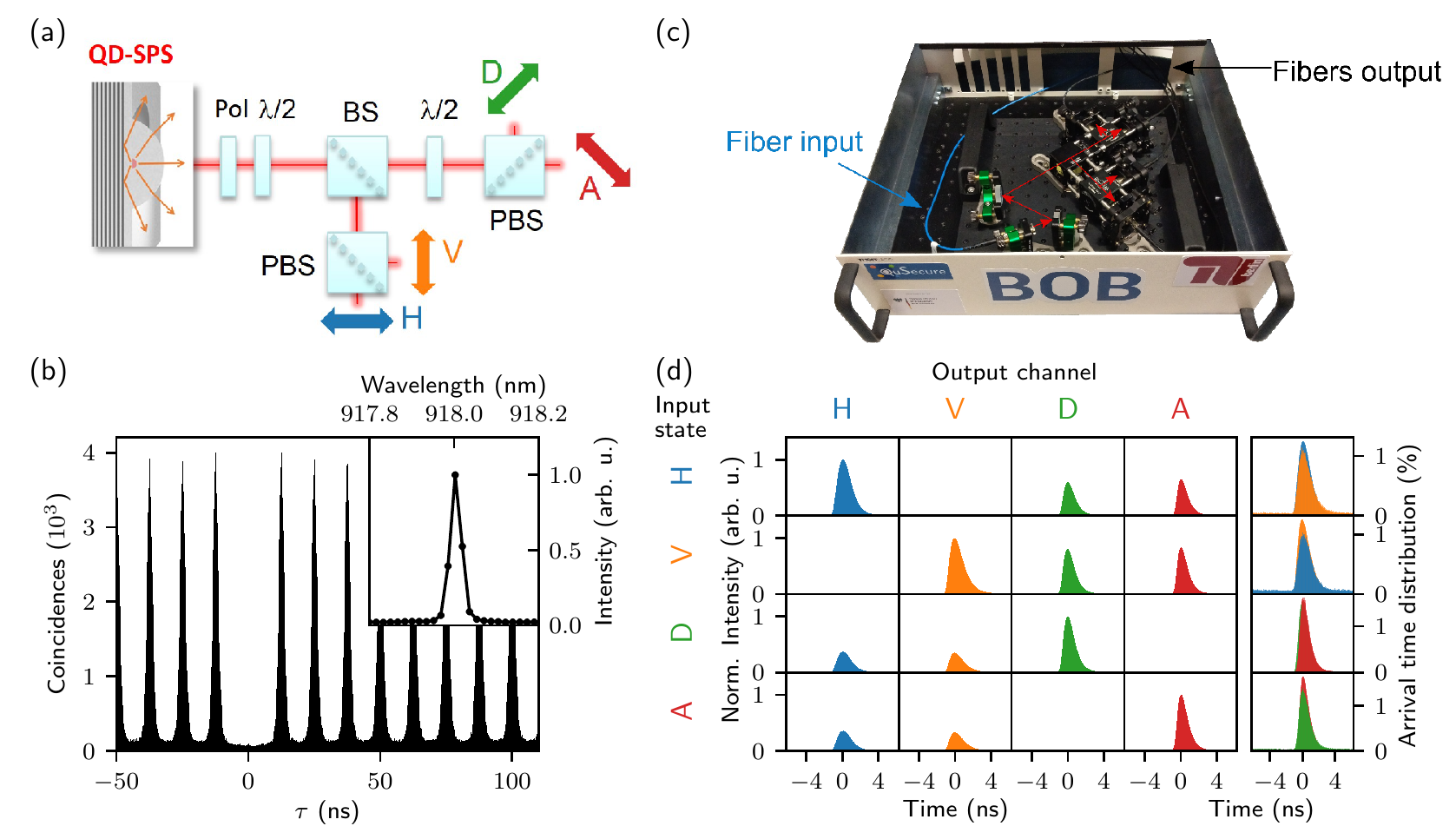}
	\caption{BB84-QKD testbed using a triggered solid-state single-photon source (SPS) and polarization coding. (a) The transmitter (Alice) sends single-photon pulses with fiexd polarization (H, V, D, and A) to the receiver module (Bob), comprising a four-state polarization analyzer. (b) Photon-autocorrelation measurement of the emission of the optically triggered SPS. Inset: Emission spectrum of the SPS, comprising a pre-selected quantum dot embedded in a photonic microlens. (c) Picture of the Bob module integrated in a 19-inch rackbox. (d) Measured photon arrival time distributions at the four detection channels of Bob for single-photon input-polarizations of H, V, D, and A. Measurement data in the left 4x4 Matrix are normalized to the maximum of the respective input state. The right panel shows the two data sets for a given input polarization basis (e.g. HH and HV) for each input state normalized to the respective for each channel photon arrival probability distributions, revealing erroneous detection events due to optical imperfections in Bob.}
	\label{fig:Setup}
\end{figure*}

The QKD testbed used for our experiments is illustrated in Fig.~\ref{fig:Setup}\,(a). 
On transmitter side, Alice is represented by a triggered SPS, comprising a single pre-selected QD embedded in a deterministically fabricated microlens \cite{Gschrey2015} providing enhanced photon collection efficiency (see Methods for details). As depicted in Fig.~\ref{fig:Setup}\,(b) this device emits single photons at an emission wavelength of 918\,nm with low multi-photon emission probability reflected in an antibunching of $g^{(2)}(0)=0.089\pm 0.002$. The non-ideal $g^{(2)}(0)$ is a consequence of the simple excitation scheme (p-shell excitation) used in our present work, and can be further improved using strict resonant pumping of the quantum emitter. The polarization state of the emitted photons is set by a high-extinction-ratio linear-film polarizer and a lambda-half waveplate, respectively, preparing single-photon pulses in horizontal (H), vertical (V), diagonal (D), and antidiagonal (A) polarization. On receiver side, Bob comprises a four-state polarization analyzer with passive basis choice. Single-photon counting modules based on silicon avalanche photon diodes, time-tagging electronics and a custom made control software is used for polarization-resolved single-photon detection, data acquisition, and postprocessing. The Bob module is integrated into a portable 19-inch rackbox presented in Fig.~\ref{fig:Setup}\,(c) (see Methods for details). In the following, we investigate the performance of this QKD testbed assuming an implementation of the BB84 protocol by analyzing the achievable QBER, single-photon purity $g^{(2)}(0)$ and secret key rate. 

First we investigate the limit our Bob module introduces to the total QBER expected in a full implementation. 
For this purpose, we record the photon arrival time distribution at the four detection channels of Bob for all four possible input polarizations of the SPS. 
The corresponding experimental data are shown in Fig.~\ref{fig:Setup}\,(d) in a 4x4 matrix representation, where the distributions within one row are normalized to the peak maximum of the curve in the respective diagonal element. Ideally, for a given input polarization (e.g. H) of one basis (H-V), one would expect only detection events in the respective channel at Bob's side (H), while the channel with orthogonal polarization (V) should be empty. Detection events in the other basis (D-A) should be equally distributed, due to the statistically random projection of the photons polarization. From the measured matrix in Fig.~\ref{fig:Setup}\,(d) this appears to be well reproduced in the experiment. A closer look in Fig.~\ref{fig:Setup}\,(d) (right panel), however, reveals the presence of erroneous detection events, by displaying the arrival time probability distributions for both polarizations of the target basis each normalized to the number of events in the given channel. In this representation, contributions of noise and optical imperfections can already be qualitatively distinguished. Correlated events in the wrong channel originate from state discrimination imperfections caused by optical imperfections of Bob (e.g. finite extinction ratios of polarizing beamsplitters and retardance deviations of waveplates), while uncorrelated background events stem from detector dark counts. 
The resulting QBER$_{\rm{Bob}}$ reads
\begin{equation}
\text{QBER}_{\text{Bob}} = \underbrace{\frac{q p_{\text{signal}}}{p_{\text{click}}}}_{\text{optical imperfections}}+\underbrace{\frac{p_{\text{dc}/2}}{p_{\text{click}}}}_{\text{dark counts}} \ , \
\label{eq:QBER_B_theo}
\end{equation}
where $q$ denotes the error contributions due to Bob's optical imperfections, $p_{\text{signal}}$ is the probability to observe a signal event, $p_{\text{dc}}$ the probability for a dark count event, and $p_{\text{click}}$ the overall probability for a click \cite{Waks2002a}. Furthermore, the distributions of photons projected in the wrong basis, i.e. D photons detected in the H channel and vice versa, are not equally distributed as ideally expected. Instead, the probability to detect an H~(V) photon in the D~(A) basis is higher compared to the case of detecting D~(A) photons in the H(V) basis. This is a result of a detection efficiency mismatch across the four detection channels, which is caused by slightly varying transmission losses in the optical paths and different quantum efficiencies of the detector modules. Please note, that the detection efficiency mismatch is important to consider in the security analysis of full implementations of QKD,  as it leads to a reduced tolerable QBER \cite{Lydersen2010}.

\subsection*{Performance optimization via temporal filtering}
\begin{figure*}[]
	\centering
	\includegraphics[width=0.95\linewidth]{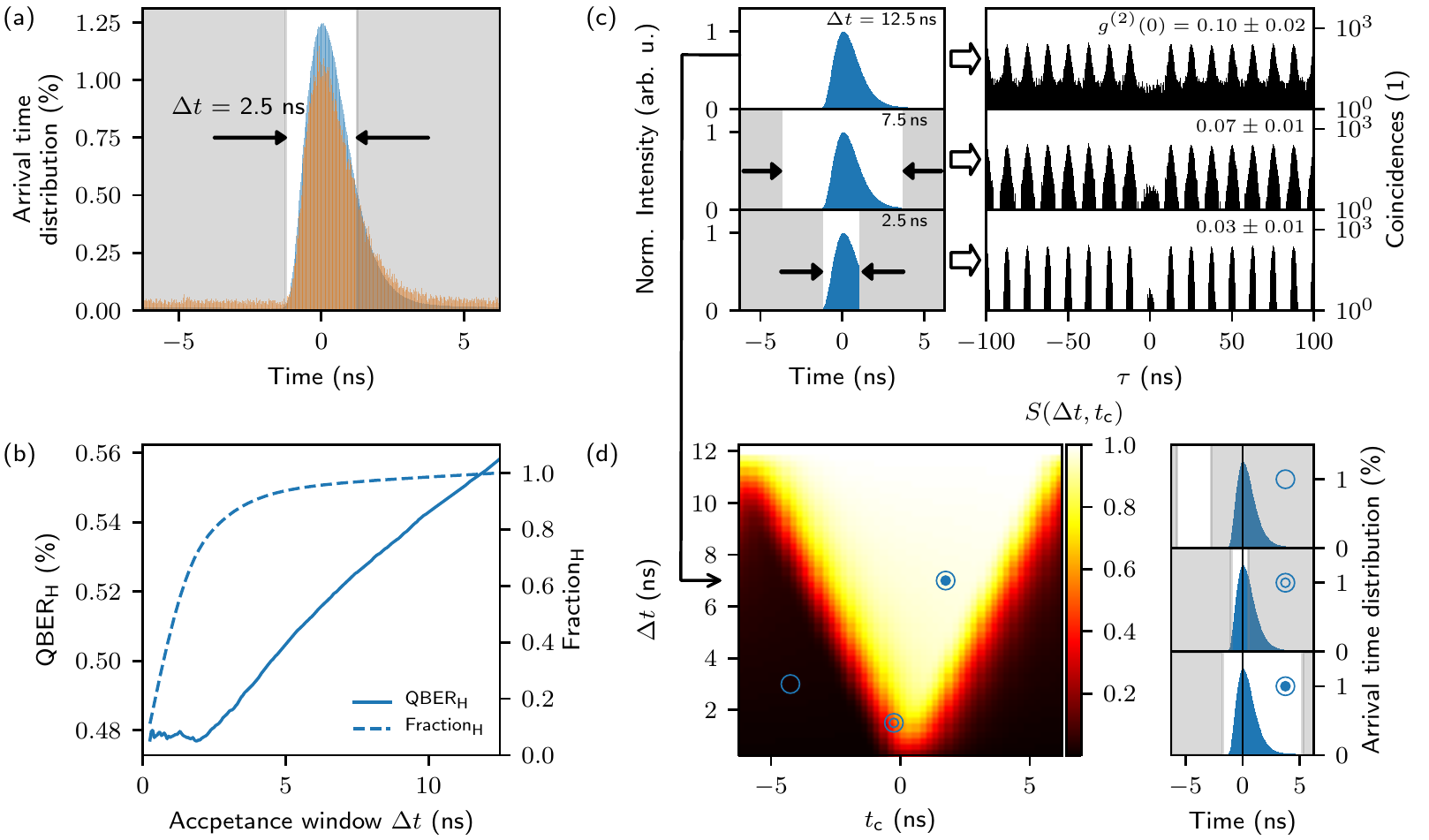}
	\caption{The effect of temporal filtering on key parameters of QKD (exemplary shown for H input polarization). (a) Photon arrival time probability distributions of H-photons detected in the H-channel (blue) and H-photons detected in the V-channel (orange). Applying temporal filtering, noise due to detector dark counts can be reduced. (b) QBER and sifted key fraction $F$ as a function of the acceptance time window width $\Delta t$ for fixed window center $t_{\text{c}}$. (c) Impact of the temporal filtering on the $g^{(2)}(0)$. Each correlation histogram $g^{(2)}(\tau)$ is calculated from the raw recorded time-tags of all four detection channels after applying the temporal filter to the data for an evaluation time of 360\,s. This evaluation time corresponds to the first data point in Fig.\ref{fig:realtime}~(d). (d)~ Expected secret key rate fraction $S(\Delta t,t_{\rm{c}})$ as a function of the temporal width $\Delta t$ and the center $t_{\text{c}}$ of the acceptance time window. For the analysis the QBER and the sifted key fraction were considered in a two-dimensional parameter space ($\Delta t$, $t_{\text{c}}$) (see Supplementary Note\,2), while $g^{(2)}(0)$ was fixed to its unfiltered value (cf. discussion in main text).}
	\label{fig:TemporalFiltering}
\end{figure*}

In the following, we analyze the impact of the temporal filtering of the raw sifted key on the performance of our single-photon QKD testbed.
Experimentally, the error contribution in the H-channel is calculated via $\text{QBER}_{\text{H}}(\Delta t, t_{\text{c}}) = N_{\text{V}}/(N_{\text{H}}+N_{\text{V}})$, where $N_{\text{H}}$ and $N_{\text{V}}$ denote the number of clicks in $H$ and $V$ polarization, respectively, detected within an acceptance time window of width $\Delta t$ centered at time $t_{\rm{c}}$. Note here, that we calculate the QBER by its definition using all events recorded in the respective acceptance time window, while it has to be carefully estimated in full implementations using subsets of bits \cite{Gao2019}.
Restricting the acceptance time window, the signal-to-noise-ratio can be enhanced, as noise due to detector dark counts can be filtered effectively \cite{Rau2014,Ko2018}.
Fig.~\ref{fig:TemporalFiltering}\,(a) exemplarily illustrates the measured photon arrival time probability distributions at both detectors of the H-V basis (H-polarized single-photon input) together with an acceptance time window $\Delta t=2.5\,$ns centered at the pulse maximum ($t_{\rm{c}}=6.25\,$ns).
Evaluating $\text{QBER}_{\text{H}}(\Delta t, t_{\text{c}})$ by applying a temporal filter to the recorded time-tags, the QBER and the fraction $F$ of the sifted and filtered photon detection events can be extracted as a function of $\Delta t$ (see Fig.~\ref{fig:TemporalFiltering}~(b)). Restricting the acceptance time window $\Delta t$ leads first of all to a reduction of the sifted key, as portions of the overall signal are discarded. At the same time the contribution of the detector dark counts is reduced, leading to a decrease of QBER$_{\rm{H}}$ towards small $\Delta t$. At $\Delta t=1.7\,$ns a local minimum with $\rm{QBER}_{\text{H}}=0.48$\% limited by optical imperfections inside the receiver is observed. This value can be further improved, e.g. by using polarization beam splitters based on Wollaston prisms enabling higher extinction ratios compared to beampslitter cubes with dielectric coating. Note, that the global minimum in $\rm{QBER}_{\text{H}}$ at $\Delta t=0.05\,$ns is not taken into account, due to the vanishing sifted key. The remaining three channels of the Bob module show similar behavior (see Supplementary Note\,1). Note, that the single-photon pulses at Bob need to be synchronized carefully for both channels of one basis to achieve optimum performance (see Methods section '˜Postprocessing').

Next, and in contrast to WCP-based implementations, the photon statistics needs to be taken into account in the security analysis for sub-Poissonian quantum light sources. The multi-photon probability $p_{\rm{m}}$ inside the quantum channel is governed by $g^{(2)}(0)$ via 
\begin{equation}
p_{\rm{m}} \leq \frac{\mu^2g^{(2)}(0)}{2}
\end{equation}
where $\mu$ is the mean photon number per pulse Alice couples to the quantum channel \cite{Waks2002a}.

In our QKD testbed we obtain the photon autocorrelation $g^{(2)}(\tau)$ by directly correlating the temporally filtered time-tags recorded at all four detection channels of Bob (see Methods). The resulting  $g^{(2)}(\tau)$ histograms are exemplary shown in Fig.~\ref{fig:TemporalFiltering}~(c) for the case of three different acceptance time windows. Narrowing the temporal filter, the antibunching improves from $0.104 \pm 0.017$ at $\Delta t = 12.50\,$ns to $0.032 \pm 0.007$ at $\Delta t = 2.5\,$ns (see Supplementary Note\,3 for further analysis). This trend is explained by the temporal filtering of two-photon emission events due to a finite probability for the re-excitation of the quantum emitter outside the acceptance time window. This effect can be used in principle to further enhance the security and the performance of QKD systems based on realistic sub-Poissonian light sources. To benefit from the temporal filtering of $g^{(2)}(0)$ at Bob, however, an active gating  (e.g. via an amplitude modulator) or at least monitoring of $g^{(2)}(0)$ would be necessary at Alice side, due to possible photon number splitting attacks outside the acceptance time window. This is an interesting perspective not considered in previous work. Due to the experimentally more demanding implementation, however, we use the $g^{(2)}(0)$ as obtained from the time-tags of the complete repetition period from now on.
Also note, that our approach of evaluating $g^{(2)}(\tau)$ is different from reports, where post-selected values of  $g^{(2)}(0)$ are generated by  the temporal filtering of  $g^{(2)}(\tau)$ after performing the correlation measurement \cite{Hanschke2018, Schoell2019}. Our routine only considers photon detection events used for secure key distillation in the end.

\begin{figure}[h!]
	\centering
	\includegraphics[width=0.95\linewidth]{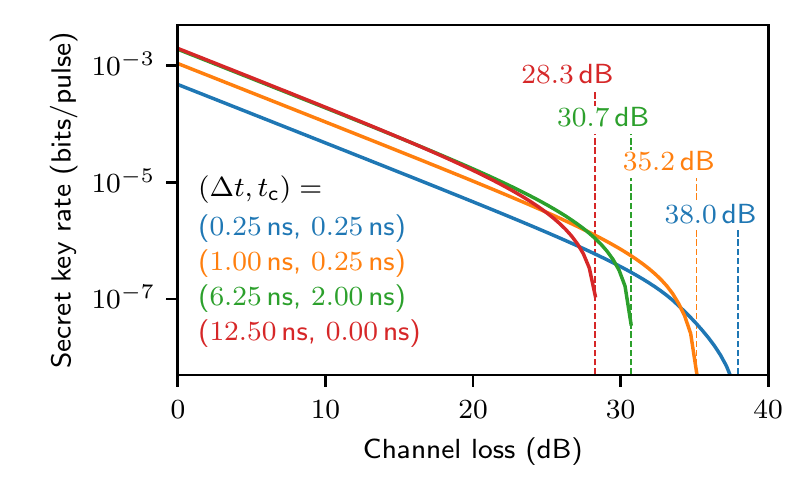}
	\caption{Optimization of single-photon QKD exploiting temporal filtering of realistic quantum light sources. Rate-loss diagram considering our experimental data from Fig.~\ref{fig:TemporalFiltering}\,(d) for different values of $\Delta t$ and $t_{\text{c}}$. Choosing optimized settings for the acceptance time window, the tolerable loss inside the quantum channel can be enhanced by 23\% in case of our SPS.}
	\label{fig:secret_key}
\end{figure}

Exploiting temporal filtering as discussed above, the overall performance of a QKD implementation based on single-photon sources can be optimized as we will demonstrate in the following.
For this purpose, a trade-off needs to be found between low QBER and high sifted key fractions on the other hand. In addition, a symmetric temporal filter as chosen above is not sufficient in general, due to the asymmetry in the photon arrival time distribution of the single-photon pulses.
To this end, we perform a two-dimensional (2D) analysis by varying the temporal width $\Delta t$ and the center $t_{\text{c}}$ of the acceptance time window.
The 2D analysis is performed for the $\text{QBER}_{\text{H}}(\Delta t,t_{\rm{c}})$ and the sifted fraction $F_{\text{H}}(\Delta t,t_{\rm{c}})$ (see Supplementary Note\,2 and 3). From these quantities, we finally extract the normalized secret key rate $S(\Delta t,t_{\rm{c}})$ expected in a full implementation of BB84 QKD according to \cite{Waks2002a}
\begin{equation}
S = \frac{p_{\text{click}}}{2}\left(\beta\tau(e)-f(e)h(e)\right) \ . \
\label{eq:ratewaks}
\end{equation}
Here, the factor $1/2$ stems from the sifting procedure, $p_{\text{click}}$ is the click-rate on the detectors, $e$ the QBER, $\beta$ the fraction of the detection events caused by single photons, $\tau(e)$ the compression function accounting for Eve's possible attacks, $h(e)$ the binary Shannon-entropy, and $f(e)$ the error correction efficiency \cite{Waks2002a}. The expected back-to-back secret key rate calculated from Eq.~\ref{eq:ratewaks} is presented in Fig.~\ref{fig:secret_key}~(a). 
A small $\Delta t$ leads to a small QBER but also to a small sifted fraction. An acceptance window within a region governed by noise does not allow for a secret key distribution at all. The optimal value for our measurement does only need to discard a small part of the signal. Depending on the length of the single-photon pulses and the detector noise level, however, temporal filtering can have crucial impact on the resulting back-to-back secure key, as demonstrated in simulations discussed further below.

The secure communication distance achievable with a given QKD system is of superior importance. Based on the secret key analysis performed in Fig.~\ref{fig:TemporalFiltering}\,(d), we calculated the rate-loss dependencies accounting for our experimental conditions (see Methods). Fig. ~\ref{fig:secret_key} illustrates the expected secret key per pulse as a function of the losses inside the quantum channel for different temporal filters. In the low-loss regime ($<20\,$dB) optimum back-to-back performance is achieved for our SPS by using the full acceptance time window ($\Delta t=12.5\,$ns), as already discussed above. The maximal tolerable loss, however, is limited to 28.28\,dB in this case. In the asymptotic case, the range could in principle be extended indefinitely by applying an asymptotic small temporal filtering and therefore enhancing the signal-to-noise ratio (SNR). In practice this is not possible. First, there is the finite temporal resolution of realistic device. The smallest possible timewindow in our system is lower bounded by the 1\,ps digitization of the times tags. Second, the reduction in sifted key by further and further narrowing the width of the acceptance time window renders the system impractical as well. The reduction to $\Delta t = 1\,$ns enhances the tolerable losses to $35.15\,$dB by decreasing the sifted fraction to 55\,\%. A further reduction down to $\Delta t = 0.25\,$ns, already below the detector timing resolution, further enhances the tolerable losses to $37.98\,$dB but reduces the sifted fraction further down to 24\,\% (see also Fig.~\ref{fig:TemporalFiltering}~(b)). Therefore, for our system by applying a reasonable temporal filter, the achievable maximal tolerable loss inside the quantum channel can be enhanced to over $35\,$dB for an optimized filter setting ($\Delta t=1.00\,$ns$, t_{\text{c}}=0.25\,$ns). This corresponds to a QKD range extension of 24\,\%. This transmission range extension is possible due to the improvement in signal-to-noise ratio resulting from the temporal filtering (see Supplementary Note\,4). This transmission range extension could be even further enhanced by exploiting the temporal filtering and monitoring of $g^{(2)}(0)$ on Alice's side, as discussed with Fig.~\ref{fig:TemporalFiltering}~(c).
Assuming a single-photon source of similar performance with an emission wavelength of 1310\,nm and 1550\,nm, respectively, extensions for the secure communication distance by 22.2\,km (to a distance of 113.4\,km) and 40.4\,km (to a distance of 206.8\,km) are expected, for state-of-the-art single-mode fiber (Corning SMF28-ULL) with 0.31\,dB/km and 0.17\,dB/km at 1310\,nm and 1550\,nm, respectively.

To assess the robustness of our SPS against transmission losses, we further analyzed its device performance in our QKD testbed in terms of the mean photon number per pulse $\mu$ (i.e. its efficiency) into the quantum channel. As derived in Ref.~\cite{Waks2002a}, a critical value $\mu_{\text{c}}=\sqrt{2p_{\text{dc}}/g^{(2)}(0)}$ can be estimated above which a sub-Poissonian light source with a given $g^{(2)}(0)$ is able to achieve the same maximally tolerable loss as the same source but with unity efficiency ($\mu$ = 1). Based on our experimental results from Fig.~2~(d) we estimated $\mu_{\text{c}}=0.0053$ for the maximal width of the acceptance time window   corresponding to a tolerable loss of $T_{\text{min}}=33.24\,$dB (see also Fig.~3). The experimental mean photon number per pulse $\mu = 0.0043$ our SPS delivers into the quantum channel (cf. Methods) therefore deviates by only $18\,\%$  from this critical value, leading to a slightly reduced maximal tolerable loss (cf. Fig.~\ref{fig:secret_key}). Also note, that the $\mu$ achieved in our QKD testbed is comparable to previous implementations of single-photon QKD by Waks et al. with $\mu = 0.007$ \cite{Waks2002}, although we do not reach the performance of more recent implementation by Takemoto et al. \cite{Takemoto2015} with $\mu = 0.050$. The considerations above show, that the requirements for maximum robustness against transmission losses can be fulfilled by relatively modest improvements of our source efficiency. To outperform WCP-based based implementations with sub-Poissonian light sources, however, a simple estimation leads to the requirement $\mu > 0.3$ (see Methods for details). The value $\mu > 0.3$ thereby is a pessimistic upper bound, which is further reduced if the QBER is not negligible. Also, finding tighter bounds for the multi-photon emission probability as a function of $g^{(2)}(0)$ will give a tighter bound on $\mu$. Recent experimental progress showed that this efficiency threshold is within reach using existing technologies using state-of-the-art deterministically fabricated solid-state quantum light sources \cite{Wang2019b}. 

Importantly, the optimization routine presented above can be adapted and extended for most other applications in quantum communication employing realistic quantum light sources, including future implementations of multi-user quantum networks based on measurement-device-independent QKD \cite{Lo2012} or multi-dimensional memory-based quantum repeaters \cite{Kuzmin2019}.

\subsection*{Real-time photon statistics monitoring}
\begin{figure}[]
	\centering
	\includegraphics[width=0.95\linewidth]{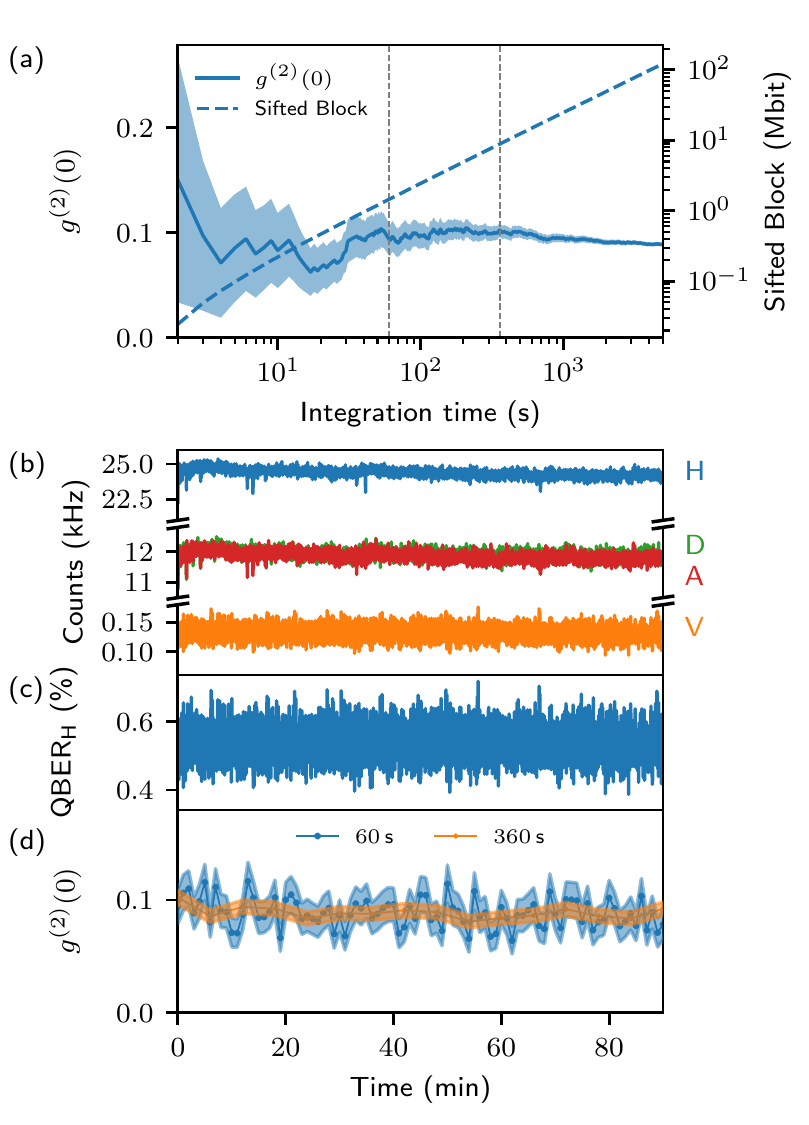}
	\caption{Real-time security monitoring for single-photon QKD. (a) Antibunching $g^{(2)}(0)$ of our SPS evaluated via correlating the time-tags of all four detection channels in non-overlapping blocks for different accumulation times compared with the corresponding length of the sifted block. (b) Clicks recorded in all four detection channels of Bob for H-polarized single-photons as input. (c) $\text{QBER}_{\text{H}}$ calculated from the data in (b). (d) Antibunching $g^{(2)}(0)$ of our SPS evaluated via correlating the time-tags of all four detection channels in non-overlapping blocks of 60\,s (blue)and 360\,s (orange) length (cf. markers in (a)).}
	\label{fig:realtime}
\end{figure}
In future QKD-secured networks implemented with quantum light sources, $g^{(2)}(0)$ inside the quantum channel needs to be monitored in real time to enable secret key distillation. Until now, most reports on single-photon QKD measured $g^{(2)}(0)$ separately and independently from the key generation/distribution process, using for instance a Hanbury-Brown and Twiss setup on Alice's side. Applying our approach for the optimization via temporal filtering presented above, we are able to also monitor $g^{(2)}(0)$ in real-time and for each block used for secret key distillation. For this purpose, we conducted a proof-of-principle experiment by recording time-tags over a period of 90\,minutes with fixed input polarization $H$ of our SPS. Based on the recorded events, we first analyze the confidence of determining $g^{(2)}(0)$ from our data. If the evaluation accumulation time is too short, the $g^{(2)}(0)$ may be over or underestimated. In the case of an overestimation this leads to reduced performance and in the case of overestimation this could lead to information leakage compromising the security. Fig.~\ref{fig:realtime}~(a) depicts the $g^{(2)}(0)$ of our SPS evaluated via Bob together with the corresponding sifted block size as a function of the accumulation time. As expected, the error decreases with increasing accumulation time. The value of $g^{(2)}(0)$ converges to $g^{(2)}(0) = 0.089\pm0.002$ for accumulation times approaching the entire measurement period. Fig.~\ref{fig:realtime}~(b) and (c) additionally show the countrates of the four detection channels and the extracted $\text{QBER}_{\text{H}}$ during the 90-min measurement period confirming stable conditions for the photon collection efficiency. Interestingly, a closer look at Fig.~\ref{fig:realtime}~(a) reveals that the extracted $g^{(2)}(0)$ does not perfectly match the converged value (within it's error) for certain ranges of accumulation times. This behavior could be related to slight changes of the properties of Alice (i.e. the SPS itself or the experimental setup) over time, which are important to consider for full implementations of quantum communication. Next, we demonstrate real-time monitoring of $g^{(2)}(0)$ inside the quantum channel evaluating the time-tags from Bob. Comparison with Fig.~\ref{fig:realtime}~(a) reveals, that for  10\,s of accumulation time the obtained $g^{(2)}(0)$ is close to the converged value but the uncertainty is far too large (43\,\%).  For 60\,s the error margin is significantly reduced to 16\,\% and the block size should already be enough to allow for a secret key distillation incorporating finite-key size effects \cite{Tomamichel2012}. Choosing 360\,s accumulation time, the uncertainty is further reduced to 6\,\% and the sifted block size would already allow secret key distillation by neglecting finite-key size effects. Fig.~\ref{fig:realtime}~(c) presents the time traces of $g^{(2)}(0)$ for two different choices of the accumulation time of 60\,s and 360\,s (cf. markers in Fig.~\ref{fig:realtime}~(a)) corresponding to sifted block sizes of 1.47, and 8.93 Mbit at the first measurement point. The real-time monitoring of $g^{(2)}(0)$ presented above, previously only been used for coherent and bunched light sources \cite{Dynes2018,Kumazawa2019}, enables us to perform a reliable security analysis in future QKD experiments, by taking into account the photon statistics of single-photon pulses used for secret key distillation. In addition, the ability to monitor the photon statistics in real time  allows for reacting on changes in the source itself or on various types of attacks if $g^{(2)}(0)$ is additionally monitored on Alice side. In case of photon number splitting attacks, for instance, an eavesdropper would artificially reduce $g^{(2)}(0)$ inside the quantum channel, which could be detected comparing $g^{(2)}_{\text{Alice}}(0)$ and $g^{(2)}(0)_{\text{Bob}}$. Moreover, any attack where the eavesdropper uses a light source with photon statistics different from the one of the QKD implementation could easily be detected.

\subsection*{Simulations}
\begin{figure}[h!]
	\centering
	\includegraphics[width=0.95\linewidth]{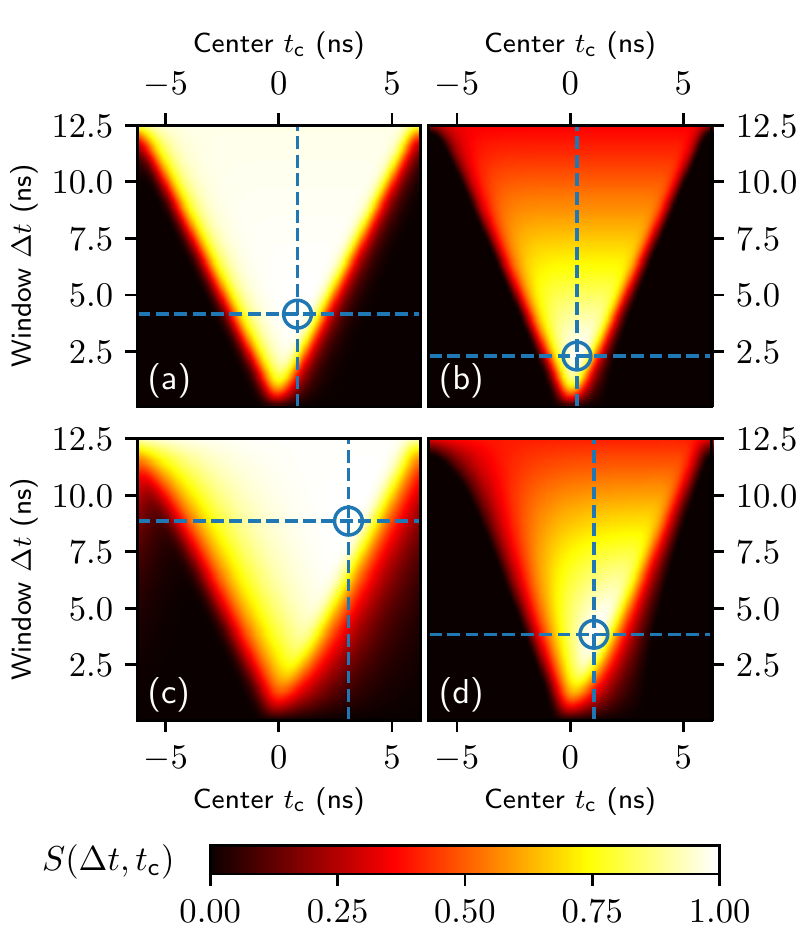}
	\caption{Simulated back-to-back secret key rates achievable via temporal filtering for different pulse lengths and noise levels: (a) Low noise and long pulse (b) high noise and short pulse (c) low noise and long pulse, as well as (d) high noise and long pulse. Careful adjustment of the acceptance time windows results in a maximum in the secret key (see markers).}
	\label{fig:simulations}
\end{figure}

To extend the scope of our approach for the performance optimization of single-photon QKD systems beyond the specific properties of our testbed, we additionally performed simulations on the secret key fraction expected for different single-photon sources and detectors. For this purpose, we modeled the photon arrival time distributions of the single-photon pulses with a synthetic pulse shape and varied the decay time constant as well as the noise level (see Methods). For the sake of clarity, we limit ourselves to four regimes: (1) Low noise and short lifetime, (2) high noise and short lifetime, (3) low noise and long lifetime, and (4) high noise and long lifetime, with short and long referring to the clock-rate. The simulation results for the secret key fraction $S(\Delta t,t_{\rm{c}})$ are presented in a two-dimensional parameter space in Fig.~\ref{fig:simulations}, assuming an ideal single-photon source ($g^{(2)}(0)=0$). 
In all four cases the temporal filtering enables one to find an optimal trade-off between sifted-key and QBER. Hence the secret key rate can be maximized by correctly choosing the settings of the temporal filter, resulting in a performance optimization of the QKD system. The gain in secret key rate compared to the case without applying a temporal filter is 2.5\,\% in case (1), 184.5\,\% in case (2), 6.0\,\% in case (3), and 148.3\,\% in case (4). Substantial improvements are achieved in the cases with high noise levels ((2) and (4)), corresponding to the regime of high transmission channel losses. Therefore, the optimization via temporal filtering becomes particular important in long distance QKD with noisy detectors. Using state-of-the-art superconducting single-photon detectors \cite{Zadeh2017}, the noise can be drastically reduced \cite{Boaron2018}. Many practical QKD scenarios, however, will not be able to provide the infrastructure for liquid-helium or closed-cycle refrigerators required for these detectors to date.

\section{Discussion}
We have demonstrated that temporal filtering of single-photon pulses is a viable tool to optimize the performance of QKD implementations based on sub-Poissonian quantum light sources. Using a basic QKD testbed comprising a solid-state based triggered SPS and a receiver module for four-state polarization coding, we showed that carefully setting the acceptance time windows enables one to maximize the achievable back-to-back secure key rate or the maximally tolerable transmission loss inside the quantum channel. Our optimization routine is particular beneficial in the high loss regime characteristic for long distance QKD. Additionally, we showed real-time security monitoring by evaluating the photon statistics of our SPS in terms of $g^{(2)}(0)$ during the key generation.

The routines developed in our work with a basic BB84-QKD testbed are readily applicable for various implementations of quantum communication employing realistic quantum light sources, including measurement device-independent QKD and quantum repeaters, and are useful for certification of QKD \cite{Tomita2019}. Furthermore, the temporal filtering and real-time monitoring of sub-Poissonian light pulses opens up new possibilities for improving the performance taking detection flaws into account. Using SPSs for QKD, an attacker is forced to use a SPS as well. This in turn reduces the penalty on the achievable secret key rate taking detection flaws into account \cite{Lydersen2010}. Even advanced non-linear attacks influencing or even controlling the photon statistics inside the quantum channel can be detected, by additionally monitoring $g^{(2)}(0)$ on Alice' side. With respect to full implementations of QKD, further extensions are required, taking side-channel attacks \cite{Rau2014,Sajeed2015} or finite-key effects \cite{Scarani2009, Curty2014, Zhang2017} into account. As the temporal filtering reduces the amount of key material that can be generated, finite-key size effects are getting increasingly important.

\section{Methods}

\subsection{Single-Photon Source}
The SPS on Alice's side comprises a single pre-selected InGaAs/GaAs QD embedded in a monolithic microlens above a bottom distributed Bragg reflector, both of which increase the photon collection efficiency from the QD. Details on the sample and its deterministic fabrication can be found elsewhere \cite{Gschrey2015,Heindel2017b}. The SPS was mounted into a closed-cycle refrigerator integrated in a cryooptical table (Model attoDRY800, attocube systems AG) for operating the SPS at a temperature of 4.2\,K. An aspheric lens ($\rm{NA}=0.77$) inside the cryostat collected the QD emission, which was optically triggered at 80\,MHz repetition rate using quasi-resonant excitation into the QD's p-shell via a pulsed (2\,ps pulse width) tunable laser system (picoEmerald, APE GmbH). Single-photon emission from the QD was spectrally filtered via an edge-pass filter and a monochromator coupled to a polarization maintaining single-mode fiber (PM 98-U25D) connected to the receiver module Bob. Here, the polarization of the single-photon pulses is set using a high-extinction-ratio linear-film polarizer followed by a lambda-half waveplate for aligning Alice's and Bob's polarization axes.

\subsection{Receiver Module Bob}
The receiver module Bob contains a four-state polarization analyzer with passive basis choice. Here, the stream of single-photon pulses is split by a non-polarizing 50:50 beamsplitter (BS) followed by a polarizing beamsplitter (PBS) in the first output and a lambda-half waveplate combined with another PBS in the second output. Thus the four BB84 states (H-, V-, D-, and A-polarized photons) are routed in four different output ports, each comprising a fiber-collimator with attached optical multimode fiber (FG050LGA, Thorlabs GmbH). The photons are detected using four single-photon counting modules (COUNT-T100-FC, Laser Components GmbH) with a timing jitter between 500\,ps and 600\,ps. The single-photon detection events are converted to four streams of time-tags (1\,ps digital resolution) using a time-to-digital converter (TDC) (quTag, qutools GmbH) synchronized to the excitation laser.

\subsection{Postprocessing}
To process the time-tags from the receiver module, a homemade software package was developed (based on LabVIEW and Rust), in order to extract the sifted key fraction, the QBER, and the antibunching value $g^{(2)}(0)$ as explained in the following. First, temporally filtered data sets were processed from the raw time-tags by discarding events outside the specified acceptance time windows of width $\Delta t$ and center $t_{\rm{c}}$. For this purpose, slight temporal delays within Bob had to be compensated using electronic delays build in the TDC electronics. This synchronization was achieved by minimizing the ratio $r = N_{\text{P}}/N_{ \overline{\text{P}}}$ of the arrival time distributions for a given polarization basis $P$ within the full temporal window of $12.5\,$ns. Note, that this temporal synchronization is important for properly extracting $g^{(2)}(\tau)$ (see further below) as well as to reduce possible detection efficiency mismatches between channels affecting the performance of QKD systems \cite{Lydersen2010}. Afterwards the parameters mentioned above were extracted from the temporally filtered data sets. The QBER was calculated from the photon arrival time distributions as well as the sifted key. The photon statistics $g^{(2)}(\tau)$ were evaluated in a $\Delta\tau=250\,$ns-wide delay window, by correlating the time-tags from the four detection channels of Bob. From the resulting $g^{(2)}(\tau)$ histograms, $g^{(2)}(0)$ was calculated via $g^{(2)}(0) = \frac{N_{\tau = 0}}{N_{\tau \neq 0}}$, where $N_{\tau = 0}$ denotes the number of coincidences of the peak at zero time delay and $N_{\tau \neq 0}$ the average number of coincidences of the side peaks. The standard error of $g^{(2)}(0)$ is deduced via Gaussian error propagation, taking into account $\sigma(N_{\tau = 0}) = \sqrt{N_{\tau = 0}}$ as well as the standard deviation of the areas from the side peaks. For illustrations in this work, a time-bin width of 25\,ps and 250\,ps were chosen for the photon arrival time distributions and $g^{(2)}(\tau)$ histograms, respectively.

\subsection{Simulations}
For the simulations the photon arrival time distributions of the single-photon pulses were modeled with synthetic pulse shapes using an exponential decay convoluted with a Gaussian of 500\,ps FWHM, accounting for the temporal response function of the detection apparatus. Two types of QD-SPSs are considered: The first one resembles an QD with a radiative lifetime of 1.5\,ns (long pulse) and the second one with a lifetime of 0.5\,ns (short pulse). The optical imperfections in the second channel were modeled by the same distribution scaled to 1\,\%. The finite signal-to-noise-ratio (noise level) was considered by an uncorrelated offset of 0.01 per bin for low noise and 0.3 per bin for high noise, corresponding to signal-to-noise ratios of 392 and 13 in the input polarization channel. To account for effects arising from the overlap of consecutive pulses, a temporal window of 12.5\,ns width was used from a train of three consecutive pulses.

\subsection{Estimation of expected secret key rates}

The expected loss-dependent secret key in Fig.~\ref{fig:secret_key} was calculated via Eq.~\ref{eq:ratewaks} using estimated parameters extracted from our measurement data with the binary Shannon entropy $h(e) = -e\log_2(e)-(1-e)\log_2(1-e)$. The parameters used for the calculation stem from the long-term measurement for fixed $H$ input polarization. The extraction from this data is described in the following for the mean photon number $\mu$, the detection rate $p_{\text{click}}$ and detector dark count probability $p_{\text{dc}}$. The mean photon number $\mu$ at Alice's output was calculated from the clock frequency of the excitation laser (\SI{80}{\mega\hertz}), the setup efficiency and the mean detector count rate on all four detectors during the measurement. This results in a mean photon number $\mu = 0.0043$. This already low value does not allow for further optimization of $\mu$ as in \cite{Waks2002a}. The detector dark counts can be estimated by shielding the detectors from all incoming light results in a cumulative dark count rate of below \SI{100}{\hertz}. For the acceptance window of \SI{12.5}{\nano\second}, this leads to $p_{\text{dc}} = 1.22\cdot10^{-6}$.

\subsection{Comparison of Device Performance }
While state-of-the-art point to point QKD systems still employ weak coherent pulses (WCPs) or even light emitting diodes \cite{Xia2019}, the performance of QKD can in principle be further enhanced by using sub-Poissonian quantum light sources. In the following we estimate the effect of different photon sources (i.e. photon statistics) on the achievable secret key rate for a given QKD system. One can show, that for the case of an ideal implementation without errors nor noise and therefore neglecting the multi-photon emission events, the secret key rate $S$ from Eq.~(\ref{eq:ratewaks}) simplifies for SPSs and for decoy-states alike to:
\begin{equation}
S_{\text{ideal}} = \eta_{\text{sifting}}f_{\text{rep}}TP(n=1).
\end{equation}
Here, $\eta_{\text{sifting}}$ denotes the efficiency of the protocol,  $f_{\text{rep}}$ the repetition rate, $T$ the transmission, and $P(n=1)$ the probability for single photon emission. Note here, that $\eta_{\text{sifting}}$ is $1/2$ in case of BB84, but this efficiency can also become close to unity by choosing asymmetrical measurement bases  \cite{Lo2004}. For a given implementation of QKD, i.e. with the same $\eta_{\text{sifting}}$, $f_{\text{rep}}$, and $T$, the performance is ultimately bounded by the probability of single photon emission $P(n=1)$ of Alice. Only single photons can be used for the secret key generation. Weak laser pulses following a Poisson photon number distribution are limited to $P(n = 1)_{\text{max}}^{\text{decoy}} \leq 0.37$ with $\mu_{\text{WCP}} = 1$, whereby this case even ignores multi-photon events. Typical WCP experiments using mean photon numbers $\mu_{\text{WCP}} = 0.5$ \cite{Lo2005} have an even lower value of  $P(n = 1) \simeq 0.3$. The upper bound for the multi-photon emission probability from \cite{Waks2002a} yields $P(n = 1) \geq \mu-\mu^2g^{(2)}(0)$. Therefore, to surpass the performance of a WCP-based QKD system at same $\eta_{\text{sifting}}$ and $f_{\text{rep}}$, Alice using an ideal SPS ($g^{(2)}(0) = 0$) must achieve $\mu = 0.3$ into the quantum channel. This efficiency is within reach using existing technologies and state-of-the-art deterministically fabricated solid-state quantum light sources \cite{Wang2019b}.
   
\section{Data availability}
The data that support the plots within this paper and other findings of this study are available from the corresponding author upon reasonable request.

\section{Acknowledgments}
We acknowledge financial support from the German Federal Ministry of Education and Research (BMBF) via the project QuSecure (Grant No. 13N14876) within the funding program Photonic Research Germany and the German Research Foundation (DFG) via SFB 787 'Semiconductor Nanophotonics: Materials, Models, Devices'.

\section{Author contributions}
T.K. designed and built the receiver module and the software used for the experiments. T.K. and M.v.H. ran the single-photon source, which was grown by J.-H.S. and A.S. and processed by M.G. and S.R. under supervision of S.R.. T.K. performed the experiments and analyzed the data, with input of L.R. and T.H.. T.K. and T.H. wrote the manuscript with input from all authors. T.H. conceived the experiment and supervised the project.

\section{Additional information}

\subsection{Competing interests}
The authors declare no competing interests.

% Bibliography
\bibliographystyle{apsrev4-1}
\bibliography{bibliography}

\clearpage
\onecolumngrid

\renewcommand{\thepage}{S\arabic{page}} 
\renewcommand{\thesection}{S\arabic{section}}  
\renewcommand{\thetable}{S\arabic{table}}  
\renewcommand{\thefigure}{S\arabic{figure}}
\setcounter{figure}{0} 
\setcounter{page}{1}

\section{Supplement}
\subsection{Note\,1: Measurement data for QBER and signal fraction for each polarization}
\begin{figure}[h]
	\centering
	\includegraphics[width=0.95\linewidth]{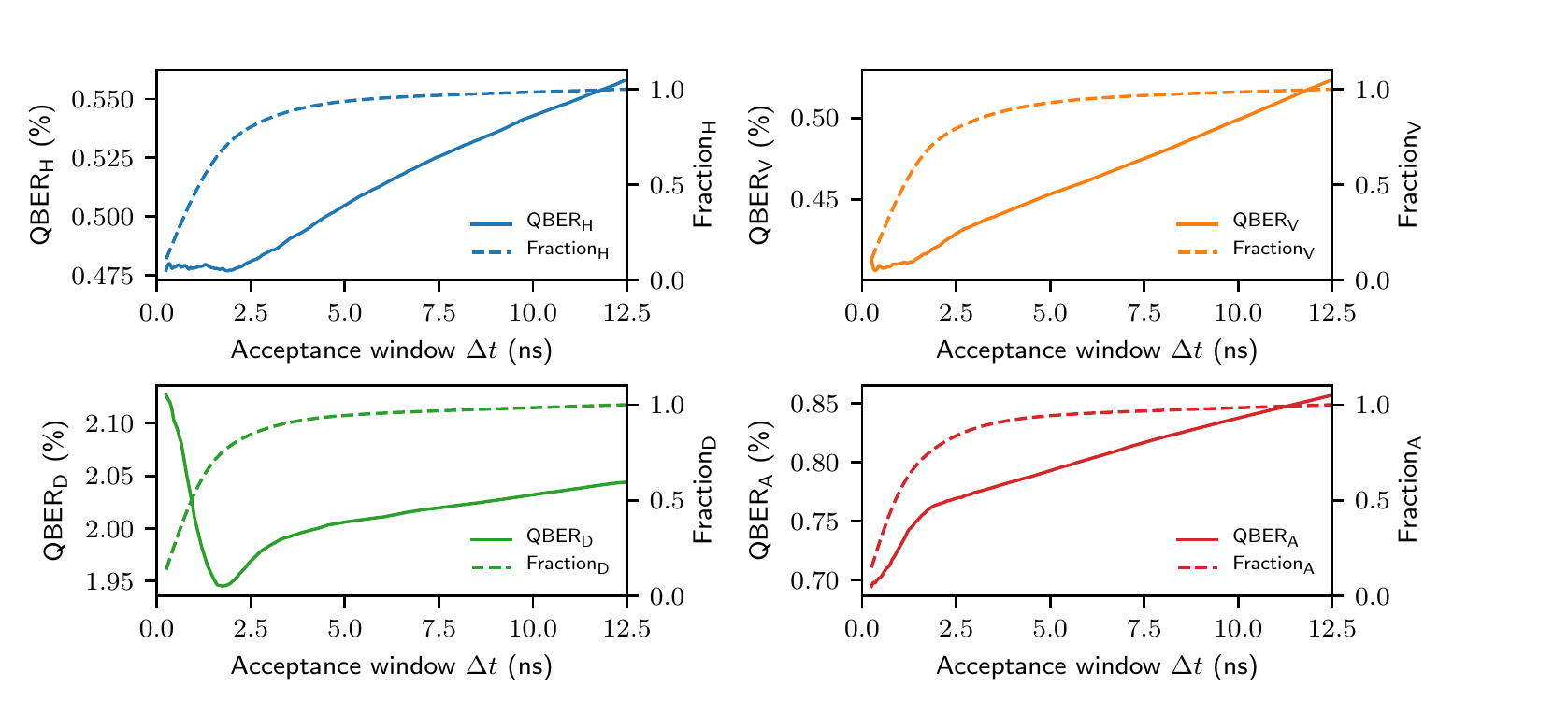}
	\caption{Experimental data complementing Fig.~2~(b) of the main article. Shown is the sifted key fraction and the QBER as a function of the width $\Delta t$ of the acceptance time window for each detection channel of Bob.}
	\label{fig:TemporalFilteringS}
\end{figure}

Fig.~\ref{fig:TemporalFilteringS} shows the sifted key fraction and the QBER extracted from the data in Fig. 1(d) of the main article for all four detection channels of Bob as a function of the width $\Delta t$ of the acceptance time window (fixed center $t_{\text c}=0\,ns$). Reducing the size of the acceptance time window, we expect a decrease in QBER due to the improvement in signal-to-noise ratio. This is observed in qualitative agreement for the H, V and A polarization, where the minimum is observed at the narrowest acceptance time window. For the D-channel we observe a differing behavior. The minimal QBER is observed at 1.75\,ns followed by a rapid increase towards diminishing width of the acceptance time window. In addition, the QBER is in general higher compared to the three remaining detection channels. These discrepancies are attributed to varying properties of the used single-photon counting modules (e.g. temporal jitter) in combination with a lower extinction ratio for the photons reflected at the polarizing beamsplitter in the D/A-Basis.

\subsection{Note\,2: Impact of 2D temporal filtering on the performance of single-photon QKD}
\begin{figure}[h!]
	\centering
	\includegraphics[width=0.95\linewidth]{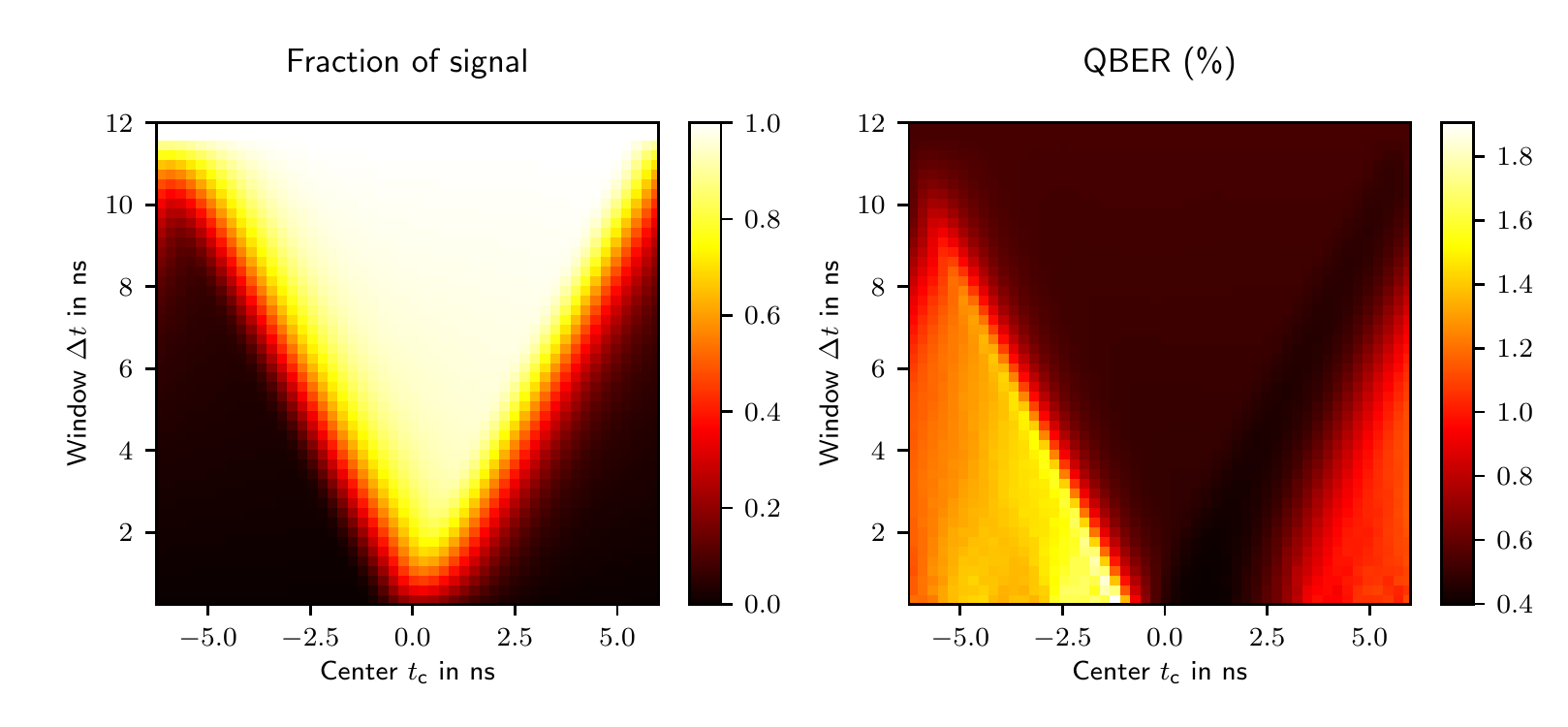}
	\caption{Experimental data complementing Fig.~2~(d) of the main article. Displayed is the sifted key fraction and the QBER as a function of the width $\Delta t$ and the center $t_{\text c}$ of the 2D acceptance time window. The smallest width is $\Delta t=0.25\,$ns for both contour plots.}
	\label{fig:2D_filtering}
\end{figure}
For signal pulses which are asymmetric in the temporal domain, the acceptance time-windows for QKD need to be adjusted in two dimensions using its width $\Delta t$ and its center position $t_{\text{c}}$. In Fig.~2~(d) of the main text, we present the secret key rate $S(\Delta t,t_{\rm{c}})$ in this 2D parameter space. Supplementary Fig.~\ref{fig:2D_filtering} depicts the complementing data for the signal fraction and the QBER.

\subsection{Note\,3: Impact of temporal filtering on the achievable communication distance}
\begin{figure}[h!]
	\centering
	\includegraphics[width=0.95\linewidth]{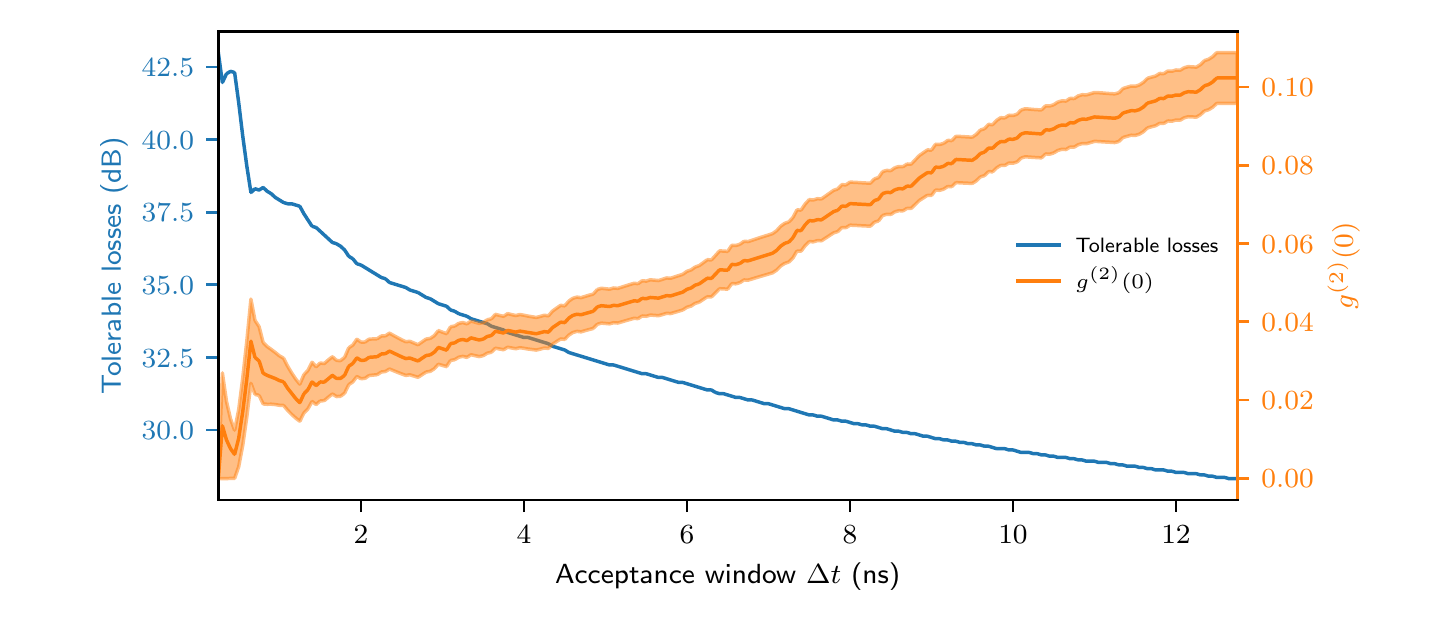}
	\caption{Experimental data complementing Fig.~2~(c) and 3 of the main article presenting the tolerable loss, the QBER$_{\text H}$ and $g^{(2)}(0)$ together with its error as a function of the width $\Delta t$ of the acceptance time window ($t_{\text c}=0$). Reducing $\Delta t$ results in an enhanced signal-to-noise ratio as indicated by the decreasing QBER and $g^{(2)}(0)$. To benefit from the temporal filtering of $g^{(2)(0)}$, active gating on Alice's side is required.}
	\label{fig:1D_filtering}
\end{figure}
As mentioned in the discussion of Fig.~3 of the main article, temporal filtering can be exploited to enhance the signal-to-noise ratio resulting in enhanced tolerable losses in single-photon QKD. Fig.~\ref{fig:1D_filtering} depicts the tolerable losses and $g^{(2)}(0)$ as a function of the width $\Delta t$ of the acceptance time window (at fixed center position $\Delta t = 0$\,ns). Exploiting temporal filtering of $g^{(2)(0)}$, the tolerable loss can be further increased compared to Fig.~3 of the main article. To benefit from this effect requires active gating on Alice's side using e.g. a fast amplitude modulator. Otherwise a backdoor is opened for photon number splitting attacks.

\end{document}